\documentstyle[preprint,aps]{revtex}
\begin{document}

\title{Black Hole Formation by Sine-Gordon Solitons
in Two-dimensional  Dilaton Gravity}
\author{Hak-Soo Shin\footnote{E-Mail:hsshin@ibm3090.snu.ac.kr}
and Kwang-Sup Soh\footnote{E-Mail:kssoh@ibm3090.snu.ac.kr}}
\address{Department of Physics Education,
Seoul National University,\\ Seoul 151-742, Korea}
\maketitle
\vspace{4cm}
\begin{abstract}
The CGHS model of two-dimensional dilaton gravity coupled to a sine-Gordon
matter field is considered. The theory is exactly solvable classically, and
the solutions of a kink and two-kink type solitons are studied in connection
with black hole formation.
\end{abstract}
\section{Introduction}
The two-dimensional dilaton gravity coupled with scalar matter fields
proposed by Callan, Giddings, Harvey and Strominger(CGHS)\cite{cghs}
has been extensivly studied with the aim of gleaning useful information
about the black hole formation and evaporation. In spite of the initial
high hope the theory turned out to be intractable even in semi-classical
approximations\cite{h3}\cite{rst} \cite{mk1} let alone in full quantum
analysis.

In the CGHS model\cite{cghs} the scalar matter was free fields, and a black
hole was formed by a shock wave of the free scalar fields. Black hole
formation by an interacting scalar field has not been considered.
On the other hand, in two-dimensional
spacetime there has been much interest in the integrable models
of the nonlinear partial differential equations, and especially in the soliton
solutions\cite{j}.
In particular sine-Gordon theory of the interacting scalar fields provides a
good example of solitons and their scatterings\cite{akns}.

In this paper we consider a sine-Gordon type matter field coupled with a
dilaton gravity, and investigate black hole formation by the solitons.
In section II we introduce the model by giving the action and gauge
fixing, and in section III we study black hole geometry formed by a kink-type
soliton. Scattering of two solitons are considered in section IV,
and brief discussions are given in the last section.

\section{Action and  Gauge Fixing}
We begin with the action in two spacetime dimensions:
\begin{equation}
S=\frac{1}{2\pi}\int_M d^2 x \sqrt{-g}\left[e^{-2\phi}[R+4(\nabla \phi)^2 +
4\lambda^2]-
\frac{1}{2}(\nabla f)^2+4\mu^2(\cos f-1) e^{-2\phi}\right], \label{eq1}
\end{equation}
where $g$, $\phi$ and $f$ are metric, dilaton, and matter fields,
respectively, and $\lambda^2$ is a cosmological constant. This action is the
CGHS
action except the last sine-Gordon term which we added to study
formation of black holes by solitons.

The classical theory described by (\ref{eq1}) is most easily analyzed in the
conformal gauge:
\begin{equation}
ds^2 = -e^{2\rho}dx^+dx^-, \label{eq2}
\end{equation}
where $x^{\pm}=t \pm x$.
The action then reduces to
\begin{equation}
S = \frac{1}{\pi} \int d^2x \left[e^{-2\phi}
(-2\partial_+\partial_-\rho +4\partial_+\phi\partial_-\phi
+\lambda^2 e^{2\rho})+\frac{1}{2}\partial_+f\partial_-f+
\mu^2(\cos f-1)e^{2\rho-2\phi} \right]
,  \label{eq3}
\end{equation}
and the metric equations of motion are
\begin{eqnarray}
T_{++} &=& e^{-2\phi} (4\partial_+\rho\partial_+\phi -2\partial_+^2\phi)+
\frac{1}{2}(\partial_+f)^2 =0, \label{eq4}
\\
T_{--} &=&e^{-2\phi}(4\partial_-\rho\partial_-\phi -2\partial_-^2\phi)
+ \frac{1}{2}(\partial_-f)^2 =0,   \label{eq5}
\\
T_{+-}&=& e^{-2\phi}(2\partial_+\partial_-\phi -
4\partial_+\phi\partial_-\phi-\lambda^2e^{2\rho})
-\mu^2(\cos f-1)e^{2\rho-2\phi} =0. \label{eq6}
\end{eqnarray}
The dilaton and matter equations are
\begin{eqnarray}
(-4\partial_+\partial_-\phi &+&4\partial_+\phi\partial_-\phi +
2\partial_+\partial_-\rho)e^{-2\phi}
+(\lambda^2 +\mu^2(\cos f-1))e^{2\rho-2\phi}=0,  \label{eq7}    \\
\partial_+\partial_-f &+&\mu^2\sin f e^{2(\rho-\phi)}=0.   \label{eq8}
\end{eqnarray}

Adding the equations (\ref{eq5}) and (\ref{eq7}) we get
\begin{equation}
\partial_+\partial_-(\rho-\phi)=0, \label{eq9}
\end{equation}
which has the general solution
\begin{equation}
\rho-\phi = w_+(x^+) +w_-(x^-). \label{eq10}
\end{equation}
The arbitrary functions $w_+$ and $w_-$ can be eliminated by fixing
the subconformal gauge freedom. From now on we will take the simplest
gauge fixing such that $\omega_+=\omega_-=0$.

Since  $\rho=\phi$ the field equations reduce to the following simple ones:
\begin{eqnarray}
\partial_+^2(e^{-2\phi}) &+& \frac{1}{2}(\partial_+f)^2 =0, \label{eq11} \\
\partial_-^2(e^{-2\phi})&+& \frac{1}{2}(\partial_-f)^2 =0,   \label{eq12} \\
\partial_+\partial_-(e^{-2\phi})&+&\lambda^2+\mu^2(\cos f-1)=0,
 \label{eq13} \\
\partial_+\partial_-f &+& \mu^2 \sin f =0, \label{eq14}
\end{eqnarray}
whose solutions we will consider in the rest of this paper.

\section{Kink Solution and Black Hole Formation}
The sine-Gordon equation(\ref{eq14}) is well-known to have a soliton solution
 \cite{tab}:
\begin{equation}
f(z) = 4 \tan^{-1}\exp{[2\frac{z-z_0}{\sqrt{1-v^2}}]},  \label{eq15}
\end{equation}
where
\begin{equation}
z=\mu(x +vt), \label{eq16}
\end{equation}
and  the center of the soliton is at $z_0$. This is a traveling wave of
kink type with velocity $v$.
Another soliton solution of anti-kink type is given by
\begin{equation}
f_{antikink}(z)=4 \tan^{-1} \exp[-2\frac{z-z_0}{\sqrt{1-v^2}}]. \label{eq17}
\end{equation}

In order to study black hole formation by a kink we solve the equationas (11
--13) with $f$ given by (\ref{eq15}).
First, we note that
\begin{equation}
f(z) = 4 \tan^{-1} \exp{[\gamma_+ x^+ +\gamma_- x^- -\Delta_0]} \label{eq18}
\end{equation}
where $\Delta_0 =\frac{2 z_0}{\sqrt{1-v^2}}$,
\begin{equation}
\gamma_+= \mu\sqrt{\frac{1+v}{1-v}},~~\gamma_-=
-\mu\sqrt{\frac{1-v}{1+v}}, \label{eq19}
\end{equation}
and
\begin{equation}
\cos f= \frac{1-6\exp({2\Delta-2\Delta_0})+\exp({4\Delta-4\Delta_0})}
{[1+\exp({2\Delta-2\Delta_0})]^2},   \label{eq20}
\end{equation}
where
\begin{equation}
\Delta\equiv \gamma_+ x^+ +\gamma_- x^-= \frac{2z}{\sqrt{1-v^2}}. \label{eq21}
\end{equation}

It is straightforward to show that the equation (\ref{eq13})
\begin{equation}
\partial_+\partial_- (e^{-2\phi}) = -[\lambda^2 + \mu^2(\cos f-1)]
\label{eq22}
\end{equation}
can be integrated as
\begin{equation}
e^{-2\phi}=a(x^+) +b(x^-) -\lambda^2 x^+x^-
-2 \log[1+\exp{(2\Delta - 2\Delta_0)}], \label{eq23}
\end{equation}
where $a(x^+)$ and $b(x^-)$ are arbitrary functions which are to be
determined by the constraint equations (\ref{eq11}) and (\ref{eq12}).
Inserting  the solutions (\ref{eq18}) and (\ref{eq23}) into (\ref{eq11})
we obtain
\begin{equation}
\frac{d^2 a(x^+)}{{dx^+}^2} = 0, \label{eq24}
\end{equation}
and similary
\begin{equation}
\frac{d^2b(x^-)}{{dx^-}^2} = 0, \label{eq25}
\end{equation}
which yields
\begin{equation}
e^{-2\phi} =C+ax^+ +b x^- -\lambda^2 x^+x^-
-2\log{[1+\exp{(2\Delta-2\Delta_0)}]}
, \label{eq26}
\end{equation}
where $a$, $b$, $C$ are constants. By shifting the origin appropriately
we can take $a=b=0$.
Hence the full solution of kink type is
\begin{eqnarray}
f&=&4\tan^{-1} \exp{(\gamma_+x^+ + \gamma_-x^- -\Delta_0)}, \label{eq27} \\
e^{-2\rho}&=& e^{-2\phi}= C-\lambda^2 x^+x^- -2\log{[1+\exp{2(\gamma_+x^+
+ \gamma_-x^- -\Delta_0)}]}, \label{eq28}
\end{eqnarray}
where $\gamma_+ =\mu\sqrt{\frac{1+v}{1-v}}$, $\gamma_+\gamma_- =-\mu^2$,
and $C$, $\Delta_0$ are constants. Anti-kink solution is similary obtainable
and the result is simply given by replacing $ (\gamma_+x^+ + \gamma_-x^-
-\Delta_0)$ by $-(\gamma_+x^+ + \gamma_-x^- -\Delta_0)$ in the equations
(\ref{eq27}) and (\ref{eq28}).

Geometry can be most easily analyzed dividing the spacetime into three regions
:$\Delta-\Delta_0 \ll -1$, $\Delta-\Delta_0 \simeq 0$,
and $\Delta-\Delta_0 \gg 1$. In the first region ($\Delta-\Delta_0 \ll -1$)
we ignore the exponential term in (\ref{eq28}), and we have
\begin{equation}
e^{-2\rho}\simeq -\lambda^2x^+x^-  \label{eq29}
\end{equation}
where we take the constant $C=0$.
It is simply the linear dilaton vacuum.
In the third region ($\Delta-\Delta_0 \gg 1$), we have
\begin{eqnarray}
e^{-2\rho}&\simeq& -\lambda^2 x^+x^- - 4(\gamma_+x^+
+ \gamma_-x^- -\Delta_0)  \nonumber \\
&=&4(\Delta_0 -\frac{4\mu^2}{\lambda^2})-\lambda^2(x^+
+\frac{4\gamma_-}{\lambda^2})(x^-+\frac{4\gamma_+}
{\lambda^2}), \label{eq30}
\end{eqnarray}
which is the geometry of a black hole of mass
$4\lambda(\Delta_0 -\frac{4\mu^2}{\lambda^2})$
after shifting $x^+$ by $\frac{4\gamma_-}{\lambda^2}$,
 and $x^-$ by $\frac{4\gamma_+}
{\lambda^2}$. The two solutions are joined along the soliton wave.
At the center of
the soliton ($\Delta-\Delta_0 \simeq 0$) we have
\begin{equation}
e^{-2\rho}\simeq 2(\Delta_0-\log2-\frac{2\mu^2}{\lambda^2})
-\lambda^2 (x^++ \frac{2\gamma_+}{\lambda^2})
(x^- + \frac{2\gamma_-}{\lambda^2})      \label{eq31}
\end{equation}
which is again the geometry of a black hole of mass
$ 2\lambda(\Delta_0-\log2-\frac{2\mu^2}{\lambda^2})$.

This analysis  shows that a black hole is formed following the input of
the soliton wave. The position of its apparent horizon,
$\partial_+\phi =0$, is given by
\begin{equation}
\lambda^2 x^- + \frac{4\gamma_+}{1+\exp{[-2(\Delta-\Delta_0)]}}=0,
\label{eq32}
\end{equation}
which, at the region $\Delta-\Delta_0 \gg 1$, is simply $x^-
= -4\frac{\gamma_+}{\lambda^2}$, that coincides with the event horizon of
the black hole,
and, at the center of the soliton wave, $x^-=-2\frac{\gamma_+}{\lambda^2}$.

In a relativistic soliton($v\rightarrow 1$) we obtain shock wave geometry
of the CGHS\cite{cghs} as

\begin{equation}
e^{-2\rho} \rightarrow \left\{\begin{array}{ll}
 -\lambda^2 x^+x^-, & x^+-x_0^+<0, \\
			     -\lambda^2 x^+x^- - 4\mu \sqrt{\frac{1+v}{1-v}}
			     (x^+ -x_0^+), & x^+-x_0^+>0,
\end{array} \right.
\label{eq33}
\end{equation}
which coincides with the solution of CGHS when the magnitude of the shock
wave $a$ is $4\mu\sqrt{\frac{1+v}{1-v}}$.
\section{Two-kink solution}
Kinks emerges unscathed from collision, suffering only a phase shift. A
two-kink solution that demonstrates this property was derived by
Perring and Skyrme\cite{ps}, which takes the form
\begin{equation}
f(x,t) = 4\tan^{-1}\left[\frac{v\sinh(\frac{\mu}{\sqrt{1-v^2}}x)}
{\cosh(\frac{\mu}{\sqrt{1-v^2}}vt)} \right]. \label{eq34}
\end{equation}
The limit $t\longrightarrow -\infty$ yield

\begin{equation}
 \lim_{t \rightarrow -\infty} f(x,t) =4 \tan^{-1}\left[\exp{(\frac{\mu
(x+vt-\delta)}{\sqrt{1-v^2}})}
 -\exp{(-\frac{\mu(x-vt+\delta)}{\sqrt{1-v^2}})}\right], \label{eq35}
 \end{equation}
 which represents a kink and an anti-kink that are separated very far and
approaching each other with the same speed $v$. Here $\delta$ is a phase
shift given as
\begin{equation}
 \delta= \sqrt{1-v^2}\ln \frac{1}{v}. \label{eq36}
 \end{equation}
 The kink and anti-kink  collide, emerge again, and run away from each other
 as $t\longrightarrow \infty$
\begin{equation}
 \lim_{t \rightarrow \infty} f(x,t) =4 \tan^{-1}
\left[-\exp{(-\frac{\mu(x+vt+\delta)}{\sqrt{1-v^2}})}
 +\exp{(\frac{\mu(x-vt-\delta)}{\sqrt{1-v^2}})}\right], \label{eq37}
 \end{equation}

 For the two kink solution the metric function is obtained as
 \begin{equation}
 e^{-2\rho}=C+a(x^+)+b(x^-)-\lambda^2x^+x^- -2\log[\cosh^2\beta t
 +v^2 \sinh^2\gamma x], \label{eq38}
 \end{equation}
 where
 \begin{equation}
 \beta =\frac{\mu v}{\sqrt{1-v^2}},~~~~~ \gamma =
\frac{\mu}{\sqrt{1-v^2}}. \label{eq39}
\end{equation}
By inserting this solution (\ref{eq38}) into (\ref{eq11}) and (\ref{eq12})
we find that
 \begin{equation}
 e^{-2\rho}=C + ax^+ + bx^- -\lambda^2x^+x^- - 2\log[\cosh^2\beta t
 +v^2 \sinh^2\gamma x]. \label{eq40}
 \end{equation}
Shifting the origin of the coordinates appropriately we have the full solution
associated with the two-kink soliton as
\begin{equation}
f(x,t) = 4\tan^{-1}\left[\frac{v\sinh(\frac{\mu (x-x_0)}{\sqrt{1-v^2}})}
{\cosh(\frac{(\mu v(t-t_0))}{\sqrt{1-v^2}})}\right], \label{eq41}
\end{equation}
 \begin{equation}
e^{-2\phi}= e^{-2\rho}=C -\lambda^2x^+x^-
- 2\log\left[\cosh^2(\frac{\mu v(t-t_0)}
{\sqrt{1-v^2}}
 +v^2 \sinh^2(\frac{\mu(x-x_0)}{\sqrt{1-v^2}})\right], \label{eq42}
 \end{equation}
 where we restored $x_0$ and $t_0$ which we had omitted for convenience.
 The constant $C$ is to be determined from the condition that the region
 of spacetime which is not affected by the incoming soliton is a
 linear dilaton vacuum. For this we consider the region
 \begin{equation}
\gamma(x-x_0)+\beta(t-t_0)\ll -1,~~~ -\gamma(x-x_0)+\beta(t-t_0)\ll -1,
{}~~~ \beta(t-t_0)\ll -1, \label{eq43}
\end{equation}
 in which region the metric function becomes
 \begin{eqnarray}
 e^{-2\rho}&\rightarrow& C-\lambda^2 x^+ x^-
- 2\log\left[\frac{e^{-\beta[(x^+ -x_0^+)
 +(x^- - x_0^-)]}}{4}\right] \nonumber \\
 &=&C+4\log2-2\beta(x_0^+ +x_0^-)+4\frac{\beta^2}{\lambda^2}
 -\lambda^2(x^+ - \frac{2\beta}{\lambda^2} )(x^- - \frac{2\beta}{\lambda^2}).
 \label{eq44}
 \end{eqnarray}
 In order to have a linear dilaton vacuum we take  $C$ as
 \begin{equation}
 C=4\beta t_0 -4(\log2+\frac{\beta^2}{\lambda^2}), \label{eq45}
 \end{equation}
 such that, in this region,
 \begin{equation}
 e^{-2\rho}=-\lambda^2(x^+ - \frac{2\beta}{\lambda^2} )
(x^- - \frac{2\beta}{\lambda^2}),
 \label{eq46}
 \end{equation}
 With this $C$ we have the solution of two approaching solitons into
 a region of the linear dilaton vacuum.

The singularity of the curvature is easily located by considering the
following three regions separately: ($R_1$) $\gamma(x-x_0)-\beta(t-t_0)
\ll\gamma(x-x_0)+\beta(t-t_0) \ll -1$, $\beta(t-t_0)\gg 1$, ($R_2$)
 $\gamma(x-x_0)-\beta(t-t_0)\gg 1$, $\gamma(x-x_0)-\beta(t-t_0)\ll -1$,
 $\beta(t-t_0)\gg 1$, and ($R_3$) $\gamma(x-x_0)+\beta(t-t_0)\gg
 \gamma(x-x_0)-\beta(t-t_0)\gg 1$, $\beta(t-t_0)\gg 1$.
 In these three regions we have
 \begin{equation}
e^{-2\rho}=\left\{\begin{array}{ll}
 (4(\beta t_0 -\gamma x_0)-2\log{v^2}
  -4\frac{\beta^2+\gamma^2}{\lambda^2})-
 \lambda^2(x^+ +\frac{2\gamma}{\lambda^2})
 (x^- -\frac{2\gamma}{\lambda^2}),&{\rm } (R_1), \\
 8\beta t_0 -\lambda^2(x^+ +\frac{2\gamma}{\lambda^2})
 (x^- +\frac{2\gamma}{\lambda^2}),& {\rm } (R_2), \\
 (4(\beta t_0 +\gamma x_0)-2\log{v^2}-4\frac{\beta^2+\gamma^2}{\lambda^2})
 -\lambda^2(x^+ -\frac{2\gamma}{\lambda^2})
 (x^- +\frac{2\gamma}{\lambda^2}), & {\rm } (R_3),
 \end{array}\right.
 \label{eq47}
 \end{equation}
 which represent the black hole metrics in each sector.
 The singularity of each region is obtained from the equation $e^{-2\rho}=0$,
 and we see that the effective masses of the black hole in each region
 is different from one another. There are two event horizons $x^- =-
 \frac{2\gamma}{\lambda^2}$ and $x^+= -\frac{2\gamma}{\lambda^2}$.

 \section{Discussion}
 There are other soliton solutions besides the simple ones we have
 considered in this paper\cite{akns}. In connection with integrable nonlinear
 systems, research on solitons is one of most interesting and active area.
 It will be quite fruitful to combine the knowledge on solitons with
 theories of two-dimensional gravity, and in particular with black hole
physics.

 It is the next step to consider Hawking radiation,
 conformal anomaly, and semi-classical analysis etc. We plan to deal with
 these topics in a separate article using the semi-classical models that
 have explicit analytic solutions such as those proposed by Bilal and Callan
 \cite{bc}, de Alwis\cite{a}, and Russo, Susskind, and Thorlacius\cite{rst}
\cite{t}.
\vspace{0.5cm}

\noindent
{\Large{\bf Acknowledgements}}\vspace{0.5cm} \\
This work was supported in part by the Center for Theoretical Physics(SNU),
and by the Basic Science Research Institute Program,  Ministry of Education,
1994, Project No. BSRI-94-2418.

\end{document}